\documentclass[10pt,twocolumn,letterpaper]{article}
\usepackage[most]{tcolorbox}  
\newtcolorbox{promptbox}[1]{
  colback=cyan!5!white,    
  colframe=cyan!30!black,   
  fonttitle=\bfseries,
  title=#1,
  boxrule=0.5pt,
  arc=4pt,
  left=6pt,
  right=6pt,
  top=6pt,
  bottom=6pt,
}

\usepackage{cvpr} 
\usepackage{times}
\usepackage{epsfig}
\usepackage{graphicx}
\usepackage{amsmath}
\usepackage{amssymb}
\usepackage{multirow} 
\usepackage{booktabs}

\begin{document}

\title{Investigating the Viability of Employing Multi-modal Large Language Models in the Context of Audio Deepfake Detection}
\author{Akanksha Chuchra$^{1}$, Shukesh Reddy$^{2}$, Sudeepta Mishra$^{1}$, Abhijit Das$^{2}$, Abhinav Dhall$^{3}$\\
$^{1}$ Indian Institute of Technology, Ropar, India \\ 
$^{2}$ Machine Intelligence Group, Birla Institute of Technology and Science, Pilani, Hyderabad Campus, India \\
$^{3}$ Monash University, Melbourne, Australia \\
{\tt\small abhijit.das@hyderabad.bits-pilani.ac.in, abhinav.dhall@monash.edu}\\
}


\maketitle
\thispagestyle{empty}

\begin{abstract}

While Vision-Language Models (VLMs) and Multimodal Large Language Models (MLLMs) have shown strong generalisation in detecting image and video deepfakes, their use for audio deepfake detection remains largely unexplored.
In this work, we aim to explore the potential of MLLMs for audio deepfake detection. Combining audio inputs with a range of text prompts as queries to find out the viability of MLLMs to learn robust representations across modalities for audio deepfake detection. Therefore, we attempt to explore text-aware and context-rich, question-answer based prompts with binary decisions. We hypothesise that such a feature-guided reasoning will help in facilitating deeper multimodal understanding and enable robust feature learning for audio deepfake detection. We evaluate the performance of two MLLMs, \textit{Qwen2-Audio-7B-Instruct} and \textit{SALMONN}, in two evaluation modes: (a) zero-shot and (b) fine-tuned. Our experiments demonstrate that combining audio with a multi-prompt approach could be a viable way forward for audio deepfake detection. Our experiments show that the models perform poorly without task-specific training and struggle to generalise to out-of-domain data. However, they achieve good performance on in-domain data with minimal supervision, indicating promising potential for audio deepfake detection.
\end{abstract}
\section{Introduction}

The rise of audio deepfakes has become a major concern in recent years \cite{PP1,audiodeepfakedetectionsurvey}. Audio deepfakes are artificially generated speech that closely mimics human voices. Advancements in generative speech techniques have enabled the creation of synthetic speech that is nearly indistinguishable from real human speech \cite{PP1,PP3}. These fake audio clips can be used to spread misinformation, impersonate individuals, and bypass voice-based security systems. As a result, audio deepfake detection has become an active area of research \cite{PP3, PP5,PP4}.
\begin{figure*}
    \centering    
    \includegraphics[width=1.05\textwidth]{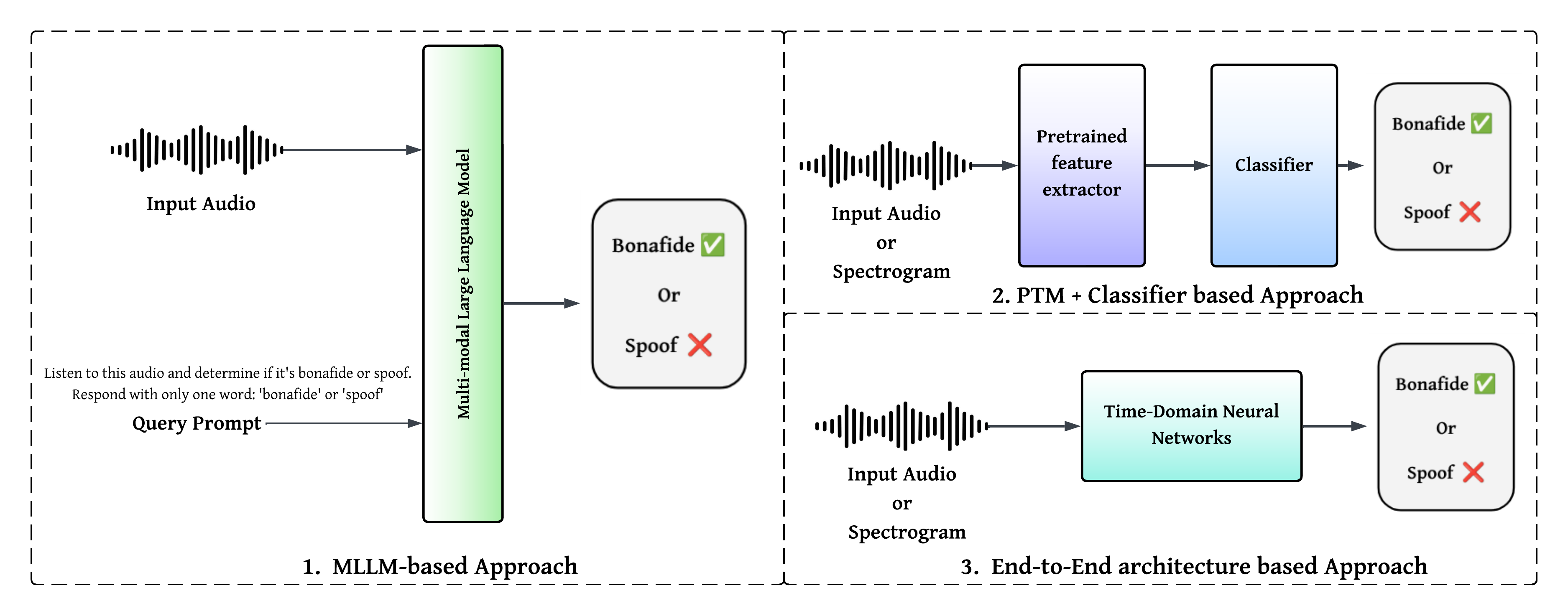}
    \caption{\textbf{Left:} Audio MLLM-based approach for audio deepfake detection, formulated as an Audio Question-Answering (AQA) task. \textbf{Right:} Traditional approaches, which rely on either end-to-end architectures or pretrained model (PTM) feature extractors followed by a classifier for discrete label prediction.}
    \label{fig:mllm_vs_traditional}
\end{figure*}

Recently, various audio deepfake detection methods such as AASIST \cite{jung2022aasist} and RawNet2 \cite{tak2021end} have been proposed that rely on end-to-end architectures, and are trained directly on the classification task. The other category of works adopted a two-stage strategy, where Pre-Trained Models (PTMs) like Whisper \cite{Whisper}, WavLM \cite{WavLM}, and wav2vec 2.0 XLS-R \cite{babu2021xls} are used as feature extractors, followed by lightweight task-specific classification heads. Interestingly, PTM-based methods often outperform end-to-end models and have shown great performance for detecting deepfakes\cite{phukan2024heterogeneity, guo2024audio, kawa2023improved, xiao2025xlsr}. This is largely attributed to the fact that these pretrained models are trained on large-scale datasets in a self-supervised manner, allowing them to learn rich and robust speech representations that generalise well across domains.


Large Language Models (LLMs) have demonstrated strong generalisation and reasoning abilities across a wide range of tasks \cite{yang2024harnessing}. Trained on diverse text corpora, they learn rich semantic representations and can follow instructions to perform tasks like text generation, question answering, and even multimodal applications with little or no task-specific training \cite{kojima2022large}.
Building upon LLMs, Multimodal Large Language Models (MLLMs) extend these capabilities further \cite{zhang2024mm}. They enable models to process and reason over multiple modalities such as text, audio, and images. This opens up new possibilities for tasks that require cross-modal understanding. Further, in recent years, Vision-Language Models (VLMs) have gained significant attention for their ability to understand and generate content across visual and textual modalities \cite{wu2024next, team2023gemini, videollama}. These models have demonstrated strong performance on a range of vision-language tasks, including image captioning, visual question answering, and multimodal reasoning \cite{zhang2024vision, ghosh2024exploring}. 

In parallel with these developments, there has been growing interest in leveraging MLLMs for media manipulation detection (such as deepfake) across different modalities, including text, images, and audio \cite{zou2025survey}. While most of the current research has focused on visual content, VLMs are increasingly being studied for their potential to identify image manipulations and detect synthetic artefacts \cite{jia2024can, yu2025unlocking}. Notably, some VLMs have also demonstrated the ability to detect visual manipulations and synthetic content in a zero-shot evaluation, i,e, without requiring additional task-specific training \cite{chang2023antifakeprompt}. The success of VLMs in detecting manipulations motivates us to study and investigate the viability of using audio-based MLLMs in the context of audio deepfake detection, which is a complete construct how it is been dealt with in the literature (See Figure~\ref{fig:mllm_vs_traditional}).

Audio MLLMs are trained on large and diverse audio-text corpora, allowing them to learn complex speech patterns and respond to instruction-like prompts. Recently, audio MLLMs have shown impressive generalisation capabilities across various speech-related tasks \cite{huang2024dynamic, yang2024air}. This work serves as one of the early efforts toward leveraging MLLMs for audio deepfake detection. The goal of our study is not only to assess whether existing audio MLLMs can perform this complex task of audio deepfake detection, but also to understand how they behave under different inference conditions and what limitations they exhibit. Thus, in this paper, we aim to address the following
questions:
\begin{itemize}
    \item Can current MLLMs be effectively utilised for the task of audio deepfake detection?
    \item How can we use MLLMs efficiently to improve audio deepfake detection in terms of feature understanding and decision accuracy? 
    \item To what extent can the MLLM-based approach enhance the generalizability of audio deepfake detection across diverse datasets and attack types? 
\end{itemize}




\section{Related Works}
The fast-paced growth of MLLMs \cite{WavLM,videollama, openai2024gpt4technicalreport,deepseekai2025deepseekv3technicalreport}, Deep Generative Models \cite{open-sora}, and Diffusion Models \cite{rombach2021highresolution} has significantly changed the domain of synthetic audio creation. The advancements in generative audio modelling have significantly reduced the obstacles to creating realistic synthetic speech, hence posing important issues of authenticity, security, and trust in audio communication. Current investigations generally adhere to the traditional pipeline model, which integrates a front-end feature extractor \cite{WavLM,Hubert,Whisper,Beats,data2vec} with a back-end classifier \cite{xception,efficient} or the end-to-end model, which directly analyses raw audio waveforms \cite{chen2024rawbmamba, liu2023leveraging}. The feature extraction, which identifies distinguishing features by detecting audio artefacts in speech signals, while end-to-end models process the audio data in its raw form to capture fine-grained details directly impacting audio deepfake detection performance.
RawNet2 \cite{tak2021end} employs Sinc-Layers to extract features directly from waveforms, while RawGAT-ST \cite{RawGAT-ST} utilises spectral and temporal sub-graphs.  Rawformer \cite{rawformer} integrates convolutional layers with Transformer architectures to represent local and global artefacts. LFCC \cite{LFCC} are widely utilised handcrafted features that employ linear filter banks, effectively capturing greater spectral information in the high-frequency domain. Nonetheless, handcrafted characteristics are compromised by biases generated due to the constraints of manual representations.  Deep features, extracted from deep neural networks, have been suggested to mitigate these constraints. Pretrained self-supervised speech models, including Wav2vec2 \cite{baevski2020wav2vec}, Hubert \cite{Hubert}, Whisper \cite{Whisper}, BEATs \cite{Beats}, WavLM \cite{WavLM}, and Data2Vec \cite{data2vec}, are the most prominent. LCNN \cite{LCNN} is a commonly used classifier, recognised as an effective baseline model in various competitions, including ASVspoof \cite{wang2020asvspoof} and ADD 2022 \cite{ADD}. 

Authors in \cite{AVSR_LLM} proposed Llama-AVSR, a multimodal large language model that executes automatic speech recognition, visual speech recognition, and audiovisual speech recognition with pretrained audio and video encoders, alongside a static large language model augmented with LoRA and lightweight projectors. 
Another work in \cite{jia2024can} leveraged GPT-4V for media forensics, focusing on video content analysis and detection of text-image misalignment. They proposed direct video processing instead of frame-level methods.
While effective, their approach is constrained by GPT-4V’s tendency to hallucinate, requiring human oversight. 

Several audio language models, including AudioGPT \cite{audiogpt}, SpeechGPT \cite{speechgpt}, LTU \cite{ltu}, Qwen2-Audio \cite{Qwen2-Audio}, DesTA \cite{Desta}, and SALMONN \cite{salmonn}, have demonstrated strong performance in tasks such as speech recognition \cite{li2024transcription,fathullah2024prompting}, audio captioning \cite{liu2023leveraging, tang2024extending,shan2025enhancing}, and audio question answering. However, their application to detecting spoofed or manipulated audio remains largely unexplored. 
To address this limitation, we investigate the capabilities of MLLMs in the context of audio deepfake detection by formulating it as an audio question-answering problem, leveraging the reasoning and perception abilities of these models.
\vspace{-5mm}
\section{Proposed Methodology}
\vspace{-2mm}
As mentioned previously, traditional audio deepfake detection uses binary classifiers optimised for discrete label prediction. In contrast, MLLMs generate predictions based on text and audio, as they are trained for next-token prediction. To align with this, with our problem of audio deepfake detection, we reformulate the task as an Audio Question-Answering (AQA) task where the model outputs either bonafide or spoof in response to the input audio and query prompt (See Figure~\ref{fig:proposed-method}). We investigate the performance of the models under two evaluation modes: the zero-shot and MLLM's fine-tuned version to enhance the task-specific performance. To ensure focused and consistent outputs, our prompts are carefully designed to instruct the model to return only the label itself, avoiding any additional explanation or comments.







\subsection{Problem Formulation}
Let $\mathcal{M} = { (\mathbf{x}_{\text{audio}}^{(i)},\ \mathbf{x}_{\text{prompt}}^{(i)}) }_{i=1}^{N}$ represent a dataset containing $N$ pairs of input audio waveforms $\mathbf{x}_{\text{audio}}^{(i)}$ and their corresponding textual prompts $\mathbf{x}_{\text{prompt}}^{(i)}$, which are processed in a multimodal pipeline. For every instance, the audio signal is passed through a pretrained audio encoder, followed by a modality adapter, to produce a sequence of audio tokens $\mathbf{A}_{\text{tokens}}$, as shown in Equation~\eqref{eq:audio_tokens}:
\begin{equation}
\mathbf{A}_{\text{tokens}} = \text{Adapter}(\text{AudioEncoder}(\mathbf{x}_{\text{audio}}))
\label{eq:audio_tokens}
\end{equation}

Simultaneously, the input prompt is processed through a tokeniser to obtain a sequence of text tokens $\mathbf{T}_{\text{tokens}}$, as defined in Equation~\eqref{eq:text_tokens}:

\begin{equation}
\mathbf{T}_{\text{tokens}} = \text{Tokenizer}(\mathbf{x}_{\text{prompt}})
\label{eq:text_tokens}
\end{equation}

\begin{figure}
    \centering
    \includegraphics[width=1\linewidth]{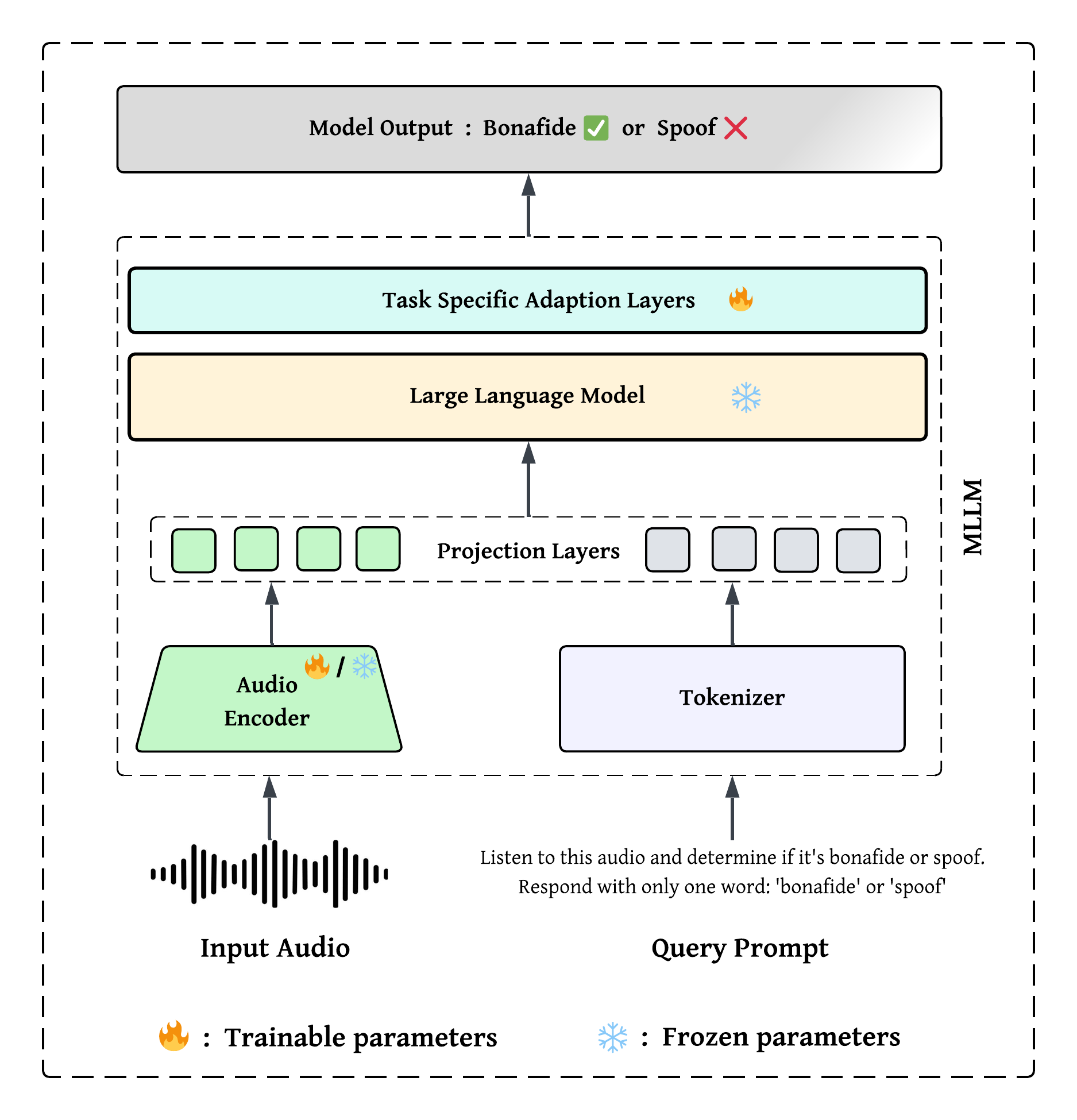}
    \caption{Overview of Proposed Audio Deepfake Detection using fine tuning of MLLMs. Fine-tuning is performed using Low-Rank Adaptation (LoRA), enabling efficient adaptation with minimal trainable parameters.}
    \label{fig:proposed-method}
\end{figure}

These two modalities, audio tokens and text tokens, are then jointly provided to the language model (See Figure~\ref{fig:proposed-method}), which generates a textual output $\mathbf{y}_{\text{out}}$ based on both sources of information, as described in Equation~\eqref{eq:llm_output}:

\begin{equation}
\mathbf{y}_{\text{out}} = \text{LLM}(\mathbf{A}_{\text{tokens}}, \mathbf{T}_{\text{tokens}})
\label{eq:llm_output}
\end{equation}
\subsection{Text Prompts}

We design and experiment with various text prompts to guide MLLMs for our task of audio deepfake detection. The text prompt serves as the input instruction to the model, shaping how it interprets the accompanying audio and generates a response. Drawing inspiration from recent work in vision-language deepfake forensics \cite{jia2024can, shahzad2024good}, our prompts are diverse in terms of question framing, instruction explicitness, and contextual richness. 

\begin{promptbox}{Prompt \#1 (Binary – Direct)}
Listen to this audio and determine if it is a bonafide or a spoof. Respond with only one word: 'bonafide' or 'spoof'.
\end{promptbox}

\begin{promptbox}{Prompt \#2 (Yes/No)}
Is this audio a spoof? Respond with 'yes' or 'no' only without any explanation.
\end{promptbox}



\begin{promptbox}{Prompt \#3 (Context-rich  Descriptive)}
Real or bonafide human speech typically has natural prosody, emotional consistency, and no robotic or glitchy artefacts. In contrast, fake or spoofed speech may exhibit issues like monotonic tone, unnatural pacing, robotic texture, or audio glitches such as clicks or distortions. Now, listen to the given audio and determine if it is bonafide or spoof based on these characteristics. Respond with only one word: 'spoof' or 'bonafide'
\end{promptbox}
\begin{promptbox}{Prompt \#Multi (Binary- Direct and Context-rich  Descriptive)}

\textbf {Binary prompt :} Listen to this audio and determine if it is a bonafide or
a spoof. Respond with only one word: 'bonafide' or
'spoof'.
\\
\textbf{Context-rich descriptive prompt:} Real or bonafide human speech typically has natural prosody, emotional consistency, and no robotic or
glitchy artefacts. In contrast, fake or spoofed speech may exhibit issues like monotonic tone, unnatural pacing, robotic texture, or audio glitches such as clicks or distortions. Now, listen to the given audio and determine if it is bonafide or spoof based on these characteristics. Respond with only one word: 'spoof' or 'bonafide'. 
\end{promptbox}
     
    
    

    

Starting with a minimal binary classification prompt, (Prompt \#1) directly asks the model to choose between "bonafide" and "spoof".  Prompt \#2 frames the question in a yes/no format, offering a slightly different linguistic structure. 
Prompt \#3 incorporates a rich contextual description, guiding the model by referencing typical auditory patterns and artefacts such as monotonic tone, robotic texture, or unnatural pacing, that are often associated with synthetic speech. Prompt \#Multi incorporates both direct binary prompt and context-rich descriptive prompt enabling the model to reason across varying levels of guidance.This progression allows us to evaluate how different levels of specificity and context influence the model's performance and response consistency. Simpler prompts are meant to reflect how a typical user might ask a straightforward question, expecting a short and direct answer. In contrast, descriptive prompts are designed to help the model reason better by including specific speech features like tone, rhythm, or glitches, that can help it decide if the audio is bonafide or spoof.

\subsection{Zero-Shot Evaluation}
In this mode, the models are evaluated directly on the audio deepfake detection task without any task-specific training or fine-tuning. The objective is to assess the inherent generalization capability of MLLMs in identifying fake speech, relying solely on their pretrained multimodal knowledge and instruction-following ability. This evaluation helps determine whether MLLMs can accurately distinguish between real and synthetic speech when guided by well-crafted prompts.

\subsection{Fine-tuning the MLLMs}
The second evaluation mode of our analysis involves fine-tuning the MLLMs using a labelled dataset tailored for the audio deepfake detection task. In this supervised setup, the models are trained on examples of both bonafide and spoofed speech, allowing them to adapt their internal representations and achieve improved performance over the zero-shot evaluation.

Fine-tuning large-scale models such as MLLMs can be computationally expensive and memory-intensive, posing significant challenges, particularly in resource-constrained settings. To mitigate this, we adopt Low-Rank Adaptation (LoRA) \cite{hu2022lora}, a parameter-efficient fine-tuning technique.
LoRA avoids the need to update the entire model by freezing the original weights and introducing small, trainable low-rank matrices into selected layers, typically within the attention and feedforward modules. This design significantly reduces the number of trainable parameters while maintaining the effectiveness of the fine-tuning . Formally, instead of updating the full weight matrix $W \in \mathbb{R}^{d \times k}$, LoRA introduces a low-rank update using matrices $A \in \mathbb{R}^{d \times r}$ and $B \in \mathbb{R}^{r \times k}$, where $r \ll \min(d, k)$. The modified weight during fine-tuning is given by:
\begin{equation}
W' = W + \Delta W = W + A B
\label{eq:lora}
\end{equation}
Here, $W$ remains frozen, and only $A$ and $B$ are optimised. After completing the fine-tuning process, we evaluate the adapted models to assess their performance, following the same evaluation setup used in the zero-shot evaluation.

\section{Experiments \& Results}
In this section we describe the experimental details of the proposed approach. Section 4.1 explains the datasets used for evaluating our models. Section 4.2 covers the implementation details and the MLLMs used in our experiments. Section 4.3 presents the results and analysis of our approach for detecting audio deepfakes.
\subsection{Dataset}
Our experiments are based on two standard datasets: the ASVspoof 2019 Logical Access dataset and the In-the-Wild (ITW) dataset. \textbf{ASVspoof 19 LA:} The ASVspoof 19 LA dataset \cite{wang2020asvspoof} is widely used in audio deepfake detection research. However, it suffers from significant class imbalance (7355 bonafide vs 63882 spoof) , with a bonafide-to-spoof ratio of roughly 1:9 across the train, development, and evaluation splits. To address this, we construct class-balanced subsets for each split referred to as \( S_{\text{train}} \), \( S_{\text{dev}} \), and \(S_{\text{eval}} \). For this, we include all available bonafide samples and randomly sampling an equal number of spoofed samples. Additionally, we ensure that the sampled spoofed audios represent all 19 attack types evenly, promoting diversity and preventing bias towards any specific attack type. We use ASV19 as a shorthand notation to refer to the full ASVspoof 2019 LA dataset. \textbf{ITW:} The In-the-Wild (ITW) dataset \cite{muller2022does} contains 31,779 samples, with 19,963 bonafide and 11,816 spoofed audios. Since the class imbalance here is less severe, we use the entire dataset for evaluation. ITW provides a more realistic and challenging benchmark due to its diverse recording conditions and spoofing methods, making it ideal for assessing the generalization capability of our models.
Table~\ref{tab:dataset_stats} summarizes the number of bonafide and spoof samples used in our experiments across the ASV19 subset splits and the ITW dataset.

\begin{table}[ht]
\centering
\caption{Dataset statistics used for training, validation, and evaluation}
\label{tab:dataset_stats}
\begin{tabular}{l|c|c|c}
\hline
\textbf{Dataset} & \textbf{\#Bonafide} & \textbf{\#Spoof} & \textbf{Total} \\
\hline
$\text{S}_{\text{train}}$ & 2580 & 2580 & 5160 \\
$\text{S}_{\text{dev}}$ & 2548 & 2548 & 5096 \\
$\text{S}_{\text{eval}}$ & 7355 & 7355 & 14710 \\
ITW & 19963 & 11816 & 31779 \\
\hline
\end{tabular}
\end{table}

\subsection{Experimental Setup}

We use two recent state-of-the-art MLLMs for our evaluation: Qwen2-Audio \cite{Qwen2-Audio} and 
SALMONN \cite{salmonn}.
Qwen2-Audio comprises two main components: an audio encoder and an LLM. The audio encoder is initialised from the Whisper-large-v3 model, while the language component is based on Qwen-7B. Specifically, for our experiments, we use the Qwen2-Audio-7B-Instruct model\footnote{https://huggingface.co/Qwen/Qwen2-Audio-7B-Instruct}, which is the chat model from the Qwen2-Audio model family. To use a shorthand notation, we refer to the model as Qwen2-Audio only. 
SALMONN \cite{salmonn} is a multimodal model that connects Vicuna LLM with two audio encoders, Whisper for speech and BEATs for general audio. These encoders process the input audio and pass their outputs to a Q-Former, which combines the features and converts them into a format the LLM can understand. Particularly, we use the SALMONN-13B \footnote{https://huggingface.co/tsinghua-ee/SALMONN} variant for our experiments. SALMONN performs well on various speech and audio tasks like ASR, translation, and emotion recognition.
We choose these models for our analysis because the models demonstrate strong performance on various tasks and established benchmarks such as Dynamic-SUPERB \cite{huang2024dynamic} and AIR-Bench-Chat \cite{yang2024air}.  
 
\subsection{Implementation Details \& Evaluation Metrics}
For LoRA fine-tuning, we set the LoRA rank to 8 and the scaling factor (alpha) to 32. A dropout of 0.1 is applied, and LoRA is integrated into the query and value projection layers of the model. We fine-tune the models using supervised fine tuning for 10 epochs with a learning rate of 1e-4. All audio samples are resampled to 16 kHz. Other than resampling, no additional pre-processing is applied. The raw audio and corresponding prompt are directly fed to the models. All experiments are performed on NVIDIA-A100 GPU. 
For evaluation, we compare the model’s textual prediction directly with the ground-truth label.
Predictions that fall outside the expected format, for example, out-of-range responses, are treated as unknown values and excluded from metric computation. For evaluation, we report accuracy and macro F1-score. Accuracy measures the overall proportion of correct predictions across both classes.
Macro F1-score, on the other hand, computes the F1-score independently for each class and then averages them, giving equal weight to both bonafide and spoof classes. 

\begin{equation}
\text{mF1} = \frac{1}{C} \sum_{i=1}^{C} \text{F1}_i = \frac{1}{C} \sum_{i=1}^{C} \frac{2 \cdot \text{Prec}_i \cdot \text{Rec}_i}{\text{Prec}_i + \text{Rec}_i}
\label{eq:macro_f1}
\end{equation}

where \( C \) is the number of classes, and \( \text{F1}_i \), \( \text{Prec}_i \), and \( \text{Rec}_i \) are the F1-score, precision, and recall for class \( i \), respectively.

\subsection{Results}

In this section, we present the performance of the evaluated models under various experimental settings. We analyse their behaviour in both zero-shot and fine-tuned modes across different prompts and datasets. Table \ref{tab:merged_finetune_comparison} summarises the performance of the MLLMs in zero-shot as well as fine-tuned settings on ASVspoof 19 eval subset. \textit{Model\textsubscript{p}} denotes a model fine-tuned using a specific prompt \textit{p}. We consider three types of prompts: Direct, Descriptive, and Multi. Here, \textit{Dir} refers to Prompt\#1, \textit{Desc} refers to Prompt\#3, and \textit{Multi} indicates fine-tuning on both Direct and Descriptive prompts. To analyse the sensitivity of model performance with respect to prompt variation, we evaluate each model using multiple prompts during inference. Table 3 presents the results for ITW dataset to check cross-dataset generalisation. 

\begin{table}[ht]
\centering
\footnotesize
\renewcommand{\arraystretch}{1.4}
\setlength{\tabcolsep}{10pt}
\begin{tabular}{l|l|cc}
\hline
\multicolumn{1}{l|}{\multirow{2}{*}{\textbf{Prompt}}} & \multicolumn{1}{l|}{\multirow{2}{*}{\textbf{Zero shot evaluation}}} & \multicolumn{2}{c}{\textbf{$\text{S}_{\text{eval}}$}} \\ \cline{3-4} 
\multicolumn{1}{l|}{}                                   & \multicolumn{1}{l|}{}            & \textbf{ACC }                     & \textbf{mF1 } \\ \hline
\multicolumn{1}{l|}{\multirow{2}{*}{\textbf{Prompt\#1}}} & \multicolumn{1}{c|}{Qwen2-Audio} & \multicolumn{1}{l}{0.34}         & 0.28                     \\
\multicolumn{1}{l|}{}                                   & \multicolumn{1}{c|}{SALMONN}     & \multicolumn{1}{l}{\underline {0.46}}         & \underline{0.46}                     \\ \hline
\multicolumn{1}{l|}{\multirow{2}{*}{\textbf{Prompt\#2}}}          & \multicolumn{1}{c|}{Qwen2-Audio} & \multicolumn{1}{l}{0.36}         & 0.22                     \\
\multicolumn{1}{l|}{}                                   & \multicolumn{1}{c|}{SALMONN}     & \multicolumn{1}{l}{0.50}         & 0.33                     \\ \hline
\multicolumn{1}{l|}{\multirow{2}{*}{\textbf{Prompt\#3}}}          & \multicolumn{1}{c|}{Qwen2-Audio} & \multicolumn{1}{l}{0.43}         & 0.38                     \\
\multicolumn{1}{l|}{}                                   & \multicolumn{1}{c|}{SALMONN}     & \multicolumn{1}{l}{0.45}         & 0.32                     \\ \hline
\multicolumn{4}{c}{\textbf{Finetuned Models}}                                                                                                             \\ \hline
\multicolumn{1}{c|}{\multirow{6}{*}{\textbf{Prompt\#1}}} & \multicolumn{1}{c|}{Qwen2-Audio \textsubscript{Dir}} & 0.96                             & \multicolumn{1}{c}{0.96} \\
\multicolumn{1}{c|}{}                                   & \multicolumn{1}{c|}{Qwen2-Audio \textsubscript{Desc}} & 0.90                             & \multicolumn{1}{c}{0.48} \\
\multicolumn{1}{c|}{}                                   & \multicolumn{1}{c|}{Qwen2-Audio \textsubscript{Multi}} & 0.59                             & \multicolumn{1}{c}{0.49} \\
\multicolumn{1}{c|}{}                                   & \multicolumn{1}{c|}{SALMONN \textsubscript{Dir}}     & 0.96                             & \multicolumn{1}{c}{0.96} \\
\multicolumn{1}{c|}{}                                   & \multicolumn{1}{c|}{SALMONN \textsubscript{Desc}}     & 0.96                             & \multicolumn{1}{c}{0.96} \\
\multicolumn{1}{c|}{}                                   & \multicolumn{1}{c|}{SALMONN \textsubscript{Multi}}     & 0.97                             & \multicolumn{1}{c}{0.97} \\ \hline
\multicolumn{1}{c|}{\multirow{6}{*}{\textbf{Prompt\#3}}} & \multicolumn{1}{c|}{Qwen2-Audio \textsubscript{Dir}} & 0.95                             & \multicolumn{1}{c}{0.95} \\
\multicolumn{1}{c|}{}                                   & \multicolumn{1}{c|}{Qwen2-Audio \textsubscript{Desc}} & 0.91                             & \multicolumn{1}{c}{0.90} \\
\multicolumn{1}{c|}{}                                   & \multicolumn{1}{c|}{Qwen2-Audio \textsubscript{Multi}} & 0.58                             & \multicolumn{1}{c}{0.47} \\
\multicolumn{1}{c|}{}                                   & \multicolumn{1}{c|}{SALMONN \textsubscript{Dir}}     & 0.96                             & \multicolumn{1}{c}{0.96} \\
\multicolumn{1}{c|}{}                                   & \multicolumn{1}{c|}{SALMONN \textsubscript{Desc}}     & 0.97                             & \multicolumn{1}{c}{0.97} \\
\multicolumn{1}{c|}{}                                   & \multicolumn{1}{c|}{\textbf{SALMONN \textsubscript{Multi}}}     & \textbf{0.98}                               & \multicolumn{1}{c}{\textbf{0.98}}   \\ \hline
\end{tabular}
\vspace{1mm}
\caption{Comparison of model performance when fine-tuned with different prompts, evaluated on the $\text{S}_{\text{eval}}$ dataset. \textit{Dir}, \textit{Desc}, and \textit{Multi} denote fine-tuning with the Direct prompt, Descriptive prompt, and both prompts combined, respectively. The results are reported for both Direct and Descriptive prompts used at inference. Subscripts with the model name indicate the prompt used during fine-tuning. }
\label{tab:merged_finetune_comparison}
\end{table}



\begin{table}[ht]
\centering
\footnotesize
\renewcommand{\arraystretch}{1.4}
\setlength{\tabcolsep}{8pt}
\begin{tabular}{l|l|cc}
\hline
\multicolumn{1}{l|}{\multirow{2}{*}{\textbf{Prompt}}} & \multicolumn{1}{l|}{\multirow{2}{*}{\textbf{Zero shot evaluation}}} & \multicolumn{2}{c}{\textbf{ITW}} \\ \cline{3-4} 
\multicolumn{1}{l|}{}                                   & \multicolumn{1}{l|}{}            & \textbf{ACC }                     & \textbf{mF1 } \\ \hline
\multirow{2}{*}{\textbf{Prompt\#1}} & Qwen2-Audio& \underline{0.66} & \underline{0.54} \\
                                   & SALMONN     & 0.54 & 0.53 \\ \hline
\multirow{2}{*}{\textbf{Prompt\#2}} & Qwen2-Audio & 0.52 & 0.52 \\
                                   & SALMONN     & 0.52 & 0.51 \\ \hline
\multirow{2}{*}{\textbf{Prompt\#3}} & Qwen2-Audio & 0.51 & 0.44 \\
                                   & SALMONN     & 0.39 & 0.31 \\ \hline 
\multicolumn{4}{c}{\textbf{Finetuned Models}} \\ \hline 
\multirow{6}{*}{\textbf{Prompt\#1}} & Qwen2-Audio \textsubscript{Dir} & 0.37 & 0.27 \\ 
                                   & Qwen2-Audio \textsubscript{Desc} & 0.36 & 0.26 \\ 
                                   & Qwen2-Audio \textsubscript{Multi}& 0.59 & 0.49 \\ 
                                   & SALMONN \textsubscript{Dir}    & 0.58 & 0.57 \\ 
                                   & SALMONN \textsubscript{Desc}    & 0.57 & 0.56 \\ 
                                   & SALMONN \textsubscript{Multi}   & 0.63   & 0.59   \\ \hline 
\multirow{6}{*}{\textbf{Prompt\#3}} & Qwen2-Audio \textsubscript{Dir} & 0.37 & 0.27 \\ 
                                   & Qwen2-Audio \textsubscript{Desc} & 0.38 & 0.27 \\ 
                                   & Qwen2-Audio \textsubscript{Multi} & 0.59 & 0.49 \\
                                   & SALMONN \textsubscript{Dir}     & 0.56 & 0.54 \\ 
                                   & SALMONN \textsubscript{Desc}     & 0.59 & 0.58 \\ 
                                   & \textbf{SALMONN \textsubscript{Multi}}     & \textbf{0.66 }  & \textbf{0.62}  \\ \hline 
\end{tabular}
\vspace{1mm}
\caption{Cross-domain evaluation of pretrained and fine-tuned Qwen2-Audio and SALMONN models on the In-the-Wild (ITW) dataset to assess generalization across domains.}
\label{tab:zeroshot}
\end{table}
\begin{table}[ht]
\centering
\footnotesize
\renewcommand{\arraystretch}{1.4}
\setlength{\tabcolsep}{8pt}
\begin{tabular}{cl|cc}
\hline
\multicolumn{2}{c|}{\multirow{2}{*}{\textbf{Traditional Models}}}                                            & \multicolumn{2}{c}{\textbf{ASV-19}} \\ \cline{3-4} 
\multicolumn{2}{c|}{}                            & \textbf{ACC} & \textbf{mF1} \\ \hline
\multicolumn{2}{c|}{CNN (STFT \& LF)}            & 0.88         & 0.90          \\ \hline
\multicolumn{2}{c|}{RNN (STFT \& LF)}            & 0.92         & 0.91         \\ \hline
\multicolumn{2}{c|}{CRNN (STFT \& LF)}           & 0.88         & 0.90         \\ \hline
\multicolumn{2}{c|}{Swin T (STFT \& LF)}         & 0.84         & 0.87         \\ \hline
\multicolumn{2}{c|}{ConvNeXt-Tiny (STFT \& LF)}  & 0.88         & 0.90         \\ \hline
\multicolumn{2}{c|}{SinC-CNN (Raw audio)}        & 0.84         & 0.87         \\ \hline
\multicolumn{2}{c|}{Whisper+MLP (Raw Audio)}     & 0.85         & 0.88         \\ \hline
\multicolumn{2}{c|}{Speechbrain+MLP (Raw Audio)} & 0.77         & 0.81         \\ \hline
\multicolumn{2}{c|}{Seamless+MLP (Raw Audio)}    & 0.86         & 0.88         \\ \hline
\multicolumn{2}{c|}{Pyannote+MLP (Raw Audio)}    & 0.64         & 0.71         \\ \hline
\multicolumn{2}{c|}{\begin{tabular}[c]{@{}c@{}}Whisper, ConvNeXt-Tiny\\  (Raw Audio, STFT \& LF)\end{tabular}} & 0.86             & 0.88             \\ \hline
\multicolumn{2}{c|}{\begin{tabular}[c]{@{}c@{}}Whisper, CNN \\ (Raw Audio, STFT \& LF)\end{tabular}}           & 0.87             & 0.89             \\ \hline
\multicolumn{2}{c|}{\begin{tabular}[c]{@{}c@{}} Rawnet2 \end{tabular}}           & 0.93             & 0.92             \\ \hline
\multicolumn{2}{c|}{\begin{tabular}[c]{@{}c@{}} RawGAT-ST \end{tabular}}           & 0.97             & 0.93             \\ \hline
\multicolumn{2}{c|}{\begin{tabular}[c]{@{}c@{}} Rawformer \end{tabular}}           & \underline{0.98}            & \underline {0.99}            \\ \hline
\multicolumn{2}{c|}{\begin{tabular}[c]{@{}c@{}}\textbf{Proposed best} \\ \textbf{SALMONN
\textsubscript{Multi}}\end{tabular}}           & \textbf{0.98   }          & \textbf{0.98  }           \\ \hline

\end{tabular}
\vspace{2mm}
\caption{Comparison of proposed method with classical methods \cite{pham2025comprehensive} on the ASV19 dataset.} 
\label{tab:asv19}
\end{table}



\begin{table}[ht]
\centering
\footnotesize
\renewcommand{\arraystretch}{2}
\setlength{\tabcolsep}{8pt}
\begin{tabular}{cl|cc}
\hline
\multicolumn{2}{c|}{\multirow{2}{*}{\textbf{Traditional Models}}} & \multicolumn{2}{c}{\textbf{ITW}} \\ \cline{3-4} 
\multicolumn{2}{c|}{}          & \textbf{ACC} & \textbf{mF1} \\ \hline
\multicolumn{2}{c|}{LCNN}      & \underline {0.65 }         & \underline {0.63  }       \\ \hline
\multicolumn{2}{c|}{LCNN-LSTM} & 0.66         & 0.62         \\ \hline
\multicolumn{2}{c|}{Mesonet}   & 0.53         & 0.53         \\ \hline
\multicolumn{2}{c|}{ResNet18}  & 0.49         & 0.46         \\ \hline
\multicolumn{2}{c|}{CRNNSpoof} & 0.41         & 0.39         \\ \hline
\multicolumn{2}{c|}{RawNet2}   & 0.33         & 0.33         \\ \hline
\multicolumn{2}{c|}{RawPC}     & 0.45         & 0.43         \\ \hline
\multicolumn{2}{c|}{RawGAT-ST} & 0.37         & 0.38         \\ \hline
\multicolumn{2}{c|}{\textbf{Proposed best SALMONN \textsubscript{Multi}}} & \textbf{0.66  }      & \textbf{0.62 }        \\ \hline
\end{tabular}
\vspace{2mm}
\caption{Comparison of different traditional models with the proposed model on ITW dataset \cite{muller2022does}.}
\label{tab:itw}
\end{table}

\textbf{Performance with different prompts:} Among the different prompts, the direct prompt i.e. prompt \#1 yields an average accuracy of around 50\%, while the descriptive prompt (prompt \#3) achieves approximately 44.5\%, when averaged across all datasets and models. This lower performance may be attributed to the longer context length or increased token complexity in descriptive prompts, which current models may struggle to handle effectively.
Overall, we do not observe a consistent pattern across prompts, indicating that the models are highly sensitive to prompt phrasing. 

\textbf{Fine-tuned vs. Zero-shot Performance.}
In the zero-shot evaluation, the models demonstrate underwhelming performance, with accuracies remaining close to chance level. We observe significant performance gains for both models when fine-tuned, even with a minimal and balanced labelled dataset, particularly on the S\textsubscript{eval} dataset. This highlights the adaptability of both models and suggests that even limited supervision can substantially improve their detection capabilities. The best scores achieved by our finetuned models are \textbf{bold}, while the best zeroshot are \underline {underlined} in table \ref{tab:merged_finetune_comparison}, \ref{tab:zeroshot}.

\textbf{Model-wise Comparison:} Between the two models, SALMONN consistently outperforms Qwen2-Audio in most evlauation modes. In zero-shot scenarios, especially on S\textsubscript{eval}, SALMONN shows superior performance. Furthermore, when fine-tuned, SALMONN continues to outperform Qwen2-Audio across both evaluation prompts, default and descriptive, demonstrating its stronger generalization and adaptability.

\textbf{State-of-the-art comparison} with classical deepfake speech detection methods on the ASV19 set is presented in Table \ref{tab:asv19}, alongside results reported in \cite{pham2025comprehensive}. The compared methods include handcrafted feature-based approaches such as Short-Time Fourier Transform (STFT), Constant-Q Transform (CQT), Linear Filter (LF), as well as models like Rawformer, RawNet2, RawPC, and RawGAT-ST. Table \ref{tab:itw} shows the performance comparison with traditional models from \cite{muller2022does} on ITW dataset.
From the comparison, it can be found that 
the proposed fine-tuned versions attain performance that is comparable to, or better than, these classical methods available in the literature. The best scores achieved by our models are \textbf{bold}, while the best SOTA results are \underline {underlined}. 

\section{Discussion and Challenges}

While significant progress has been made in the domain of VLMs for deepfake forensics, the development of audio deepfake detection via MLLMs remains relatively limited. In comparison to their vision-language counterparts, the performance of audio MLLMs on speech-related tasks is quite comparable, hence worthwhile to use audio MLLMs for audio deepfake detection. But there could be many challenges which is attributed to the inherent characteristics of the audio modality itself. Audio data is inherently high-dimensional, containing dense temporal and frequency information that makes it more complex to model and interpret effectively. Unlike images and videos, which benefit from spatial structure and immediate visual interpretability, audio signals are abstract and require specialized transformations such as spectrograms or learned embeddings for meaningful analysis. 

This complexity is further compounded by the lack of large-scale, high-quality, instruction-aligned audio datasets, which restricts both the training and benchmarking of robust audio MLLMs. Hence, a similar extension for audio MLLMs could involve localisation in the time-frequency domain, identifying specific regions or features in the audio signal where spoofing artefacts are present. Advancing toward this level of interpretability could significantly improve trust and transparency in audio deepfake detection. Now, we discuss notable challenges and limitations encountered during the study, which we believe can guide and inform future research efforts in this emerging area. We proceed to revisit the key research questions outlined in the introduction and present insights drawn from our experimental analysis. Following this, we also highlight key challenges and limitations identified during the study, which we believe can inform and guide future research in this evolving field. 

\noindent \textbf{Q: Can MLLMs be effectively utilized for the task of audio deepfake detection ?}

As mentioned previously that the success of VLMs in deepfake media forensics has achieved an underwhelming performance.
These successes give rise to several important questions, such as the use of MLLMs for audio deepfake detection. Our findings indicate that when MLLMs can indeed be used to leverage audio deepfake detection. MLLMs trained with multi-prompt input demonstrated promising capabilities in identifying spoofed audio, as evident from Tables \ref{tab:merged_finetune_comparison} and \ref{tab:zeroshot}.
\\

\noindent \textbf{Q: How can we use MLLMs improve audio deepfake detection in terms of feature understanding and decision
accuracy?}

MLLMs offer a paradigm shift in how audio deepfake detection can be approached, through instruction-guided reasoning rather than static, fixed-feature classification. Instead of depending solely on handcrafted acoustic features or learned embeddings, MLLMs can also interpret natural language prompts. These prompts can direct their attention to specific audio traits, such as "robotic texture," "monotonic tone," or "auditory glitches", which are traits often associated with synthetic speech. This flexibility enables a more explainable and adaptive detection pipeline. However, our experimental findings reveal that current MLLMs still exhibit limited intrinsic understanding of deepfake-specific acoustic cues, particularly in the absence of task-specific supervision or fine-tuning. However, with task-specific finetuning and proper text prompt design, it is able to achieve improved results reported in Table \ref{tab:merged_finetune_comparison} and \ref{tab:zeroshot}.   
\\

\noindent \textbf{Q: To what extent can MLLM-based approach enhance the generalizability of audio deepfake detection across diverse datasets and attack types ?}

Our findings show that while fine-tuned MLLMs achieve strong performance on in-domain evaluation (S\textsubscript{eval}) as demonstrated in Table \ref{tab:merged_finetune_comparison}, their generalization to out-of-domain data such as the ITW dataset remains limited and comparable to reported in Table \ref{tab:itw}. Despite training on varied spoof attack types, the models often overfit to the training distribution and exhibit biased behaviour, frequently labelling most ITW samples as spoof. We assume that improving generalizability may require more than just hyperparameter tuning or scaling dataset size. Techniques like few-shot learning and prompt engineering, especially using explanatory or chain-of-thought prompts, can help models focus on robust, domain-independent audio patterns. These strategies may improve the model’s ability to adapt to unfamiliar inputs and diverse spoofing scenarios.



The limited performance of audio MLLMs on complex tasks like generalised audio deepfake detection can also be attributed to the fundamental challenge of describing audio in natural language. In VLM training, image descriptions are often semantically rich, encompassing scene elements, object attributes, emotional cues, and contextual information. These diverse and detailed annotations help the models build strong cross-modal associations. In contrast, describing audio, especially synthetic or manipulated speech, tends to be less intuitive, often lacking in vocabulary or structure that captures its subtle acoustic nuances. As a result, audio MLLMs struggle to form equally deep semantic representations, limiting their effectiveness in complex downstream tasks.

Moreover, traditional deepfake detection methods rely on binary classifiers that output class probabilities, making evaluation metrics like EER and AUC naturally applicable. However, when using MLLMs, the problem formulation shifts to an audio question-answering task. In this setting, the model generates responses conditioned on both the input audio and the prompt, producing output through token prediction rather than class probability estimation. This fundamental difference makes direct comparison with conventional classifiers less meaningful. Additionally, MLLMs are prone to hallucinations, sometimes generating responses that sound convincing but are actually incorrect or unrelated. These factors highlight the need to reconsider and potentially redesign evaluation metrics that better reflect the generative nature of MLLMs in deepfake detection tasks.


\section{Conclusion \& Future Work}
This study serves as an initial exploration into the applicability of state-of-the-art MLLMs for audio deepfake detection. We evaluate two recent models, Qwen2-Audio-7B-Instruct and SALMONN, to understand their effectiveness in identifying synthetic speech.
Our investigation spans two evaluation modes: a zero-shot evaluation, where the models are evaluated without any task-specific training, and fine-tuning, where the models are adapted using labelled data. In the zero-shot evaluation, both models struggle to reliably distinguish between bonafide and spoofed audio, with accuracy often falling near random chance levels (50\%). When fine-tuned on a minimal labeled dataset, the models show improvement in detection performance on in-domain data. However, their ability to generalize to more challenging and diverse datasets such as the In-the-Wild dataset remains limited. In the future, we aim to develop specialized MLLM-based architectures specifically tailored for audio deepfake detection. 

Beyond improving raw detection performance, we also plan to explore the use of MLLMs for explainability in audio deepfake detection. Given their ability to generate natural language outputs, these models can be leveraged to produce interpretable justifications for their predictions. This direction remains largely underexplored and could provide valuable insights into the model's decision-making process, adding to more transparency in audio deepfake detection. Furthermore, most existing work assumes the entire audio is either real or fake. A promising direction is to explore partial deepfake localization using MLLMs, where only segments are manipulated. MLLMs, with well-designed prompts, could help identify such regions and explain anomalies like glitches or unnatural prosody in human-readable terms.
\section*{Acknowledgements}
This work was funded by IDEAS, TiH, ISI, Kolkata, DST, Government of India under the project titled "Generalized Tampering Detection in Media (GTDM)" and project number OO/ISI/IDEAS-TIH/2023-24/86.
{\small
\bibliographystyle{ieeetr}
\bibliography{ref}
}

\end{document}